\documentclass[default,iicol]{sn-jnl}

\jyear{2023}%

\theoremstyle{thmstyleone}%
\theoremstyle{thmstyletwo}%

\theoremstyle{thmstylethree}%

\usepackage{lineno,hyperref}
\usepackage{multirow}
\usepackage{graphicx}
\usepackage{xcolor}
\modulolinenumbers[5]

\usepackage{longtable,booktabs,array,multirow,tabularx,ragged2e,array}
\usepackage{breqn}
\usepackage{subcaption}

\newcolumntype{L}[1]{>{\raggedright\let\newline\\\arraybackslash\hspace{0pt}}m{#1}}
\newcolumntype{C}[1]{>{\centering\let\newline\\\arraybackslash\hspace{0pt}}m{#1}}
\newcolumntype{R}[1]{>{\raggedleft\let\newline\\\arraybackslash\hspace{0pt}}m{#1}}

\usepackage{adjustbox}

\usepackage{etoolbox}
\usepackage{soul} 

\newtoggle{finalPaper}

\setstcolor{blue}

\toggletrue{finalPaper} 

\iftoggle{finalPaper} {
	\newcommand{\addtxt}[1]{#1}
	\newcommand{\change}[2]{#2}
	\newcommand{\rmvtxt}[1]{}}
{
    \soulregister\cite7 
    \soulregister\ref7
    
	\newcommand{\addtxt}[1]{\textcolor{blue}{#1}}
    \newcommand{\change}[2]{\texorpdfstring{\st{#1}\textcolor{blue}{#2}}{#2}}
    \newcommand{\rmvtxt}[1]{\texorpdfstring{\st{#1}}{#1}}
}

\newtoggle{removeItalic}
\togglefalse{removeItalic} 
\iftoggle{removeItalic} {
    \renewcommand{\textit}[1]{#1}
}

\begin{document}

\title[Studying Drowsiness Detection Performance while Driving through Scalable Machine Learning Models using Electroencephalography]{Studying Drowsiness Detection Performance while Driving through Scalable Machine Learning Models using Electroencephalography}


\author[1]{\fnm{José Manuel} \sur{Hidalgo Rogel}}\email{josemanuel.hidalgor@um.es}

\author[1]{\fnm{Enrique Tomás} \sur{Martínez Beltrán}}\email{enriquetomas@um.es}

\author[1]{\fnm{Mario} \sur{Quiles Pérez}}\email{mqp@um.es}

\author*[1]{\fnm{Sergio} \sur{López Bernal}}\email{slopez@um.es}

\author[1]{\fnm{Gregorio} \sur{Martínez Pérez}}\email{gregorio@um.es}

\author[2]{\fnm{Alberto} \sur{Huertas Celdrán}}\email{huertas@ifi.uzh.ch}

\affil*[1]{\orgdiv{Department of Information and Communications Engineering}, \orgname{University of Murcia}, \orgaddress{\street{Campus de Espinardo}, \city{Murcia}, \postcode{30100}, \state{Murcia}, \country{Spain}}}

\affil[2]{\orgdiv{Communication Systems Group CSG, Department of Informatics IfI}, \orgname{University of Zurich UZH}, \orgaddress{\city{Zürich}, \postcode{8050}, \state{Zürich}, \country{Switzerland}}}


\abstract{
\textbf{Background / Introduction} \\
Driver drowsiness is a significant concern and one of the leading causes of traffic accidents. Advances in cognitive neuroscience and computer science have enabled the detection of drivers' drowsiness using Brain-Computer Interfaces (BCIs) and Machine Learning (ML). However, the literature lacks a comprehensive evaluation of drowsiness detection performance using a heterogeneous set of ML algorithms, \change{and it is}{being also} necessary to study the performance of scalable ML models suitable for groups of subjects. 

\textbf{Methods} \\
To address these limitations, this work presents an intelligent framework employing BCIs and features based on electroencephalography for detecting drowsiness in driving scenarios. The SEED-VIG dataset is used to evaluate the best-performing models for individual subjects and groups. 

\textbf{Results} \\
Results show that Random Forest (RF) outperformed other models used in the literature, such as Support Vector Machine (SVM), with a 78\% f1-score for individual models. Regarding scalable models, RF reached a 79\% f1-score, demonstrating the effectiveness of these approaches. This publication highlights the relevance of exploring a diverse set of ML algorithms and scalable approaches suitable for groups of subjects to improve drowsiness detection systems and ultimately reduce the number of accidents caused by driver fatigue. 

\textbf{Conclusions} \\
The lessons learned from this study show that not only SVM but also other models not sufficiently explored in the literature are relevant for drowsiness detection. Additionally, scalable approaches are effective in detecting drowsiness, even when new subjects are evaluated. Thus, the proposed framework presents a novel approach for detecting drowsiness in driving scenarios using BCIs and ML.
}



 
\maketitle

\section{Introduction}\label{sec:1_intro}
Drowsiness is defined as a person's tendency to fall asleep. This situation is especially critical in driving scenarios, where the dangerous combination of driving and sleepiness commonly happens \cite{Kamran2019}. Particularly, the National Highway Traffic Safety Administration (NHTSA) reported between 2013 and 2019 a total of 5\change{~}{.}593 fatalities in motor vehicle crashes involving drowsy drivers. In 2017, exclusively in the USA, 91\change{~}{.}000 police-reported crashes involved drowsy drivers, which led to about 50\change{~}{.}000 people being injured \cite{iii-drowsy-driving}.

In the past years, drowsiness assessment has become a topic of interest for researchers. In this sense, cognitive neuroscience, the area of knowledge responsible for studying the nervous system that supports mental functions \cite{neurobiology-Shepherd}, including drowsiness, has proposed different techniques for its quantification \cite{drowsiness-assessment-techniques}. The first ones are based on monitoring subjects' behavior such as facial expressions, heart rate, and yawning in order to assess drowsiness. Although these techniques represent an advance in safety, they have significant limitations since they produce false positives and false negatives, not always being able to measure attributes related to fatigue or drowsiness.

Next, solutions based on self-assessment with scales emerged. This \change{technique}{approach} consists in asking subjects to describe how drowsy they felt in the \change{last}{previous} minutes. Examples of \change{this technique}{these tests} are the Karolinska Sleepiness Scale (KSS) \cite{KSS} and the NASA Task Load Index (NASA-TLX) \cite{NASA-TLX}. However, \change{the main drawback of these methods is the inclusion of subjectivity in their self-evaluation}{this self-evaluation process introduces a subjectivity factor that represents the main drawback of these methods}. Thus, the need to objectively quantify the sleepiness of an individual arises. For this reason, neurophysiological tests have been developed, based on monitoring \change{the patient's}{patients'} brain signals to precisely identify drowsiness.

Brain signals are commonly obtained by electroencephalography (EEG), which measures the electrical activity produced in the brain through electrodes acting as sensors \cite{bci-1020-system}. The different levels of brain activity are related to the different cognitive states of the subject. Due to this, it is necessary to study the EEG signals in different frequency bands, being the lower frequency rhythms (delta, theta, and alpha) directly related to the states of relaxation and drowsiness, and the higher rhythms (beta and gamma) related to concentration and moderate mental load, and even stressful situations in the case of the gamma band \cite{Ward:book:2019, Ramadan2021}.

Brain-Computer Interfaces (BCIs) are normally used when studying EEG, where two categories are distinguished depending on the degree of invasiveness of the electrodes. On the one hand, invasive BCIs locate the electrodes within the skull, requiring a surgical process. On the other hand, non-invasive BCIs place their electrodes directly on the subject's scalp, avoiding a surgical procedure. Nevertheless, non-invasive BCIs data must be processed afterwards to remove artifacts caused by the subjects' activity\addtxt{, such as eye blinking or body movements} \cite{bci-full,bci-team}. Due to their advantages and feasibility of experimenting with subjects, non-invasive BCIs are the ones commonly used \change{for}{in} the drowsiness detection scenario. In addition to non-invasive BCIs, Machine Learning (ML) models are also used to assess drowsiness using the data collected by the BCI. For this purpose, the BCIs acquire the brain signals when the subject is driving. Then, they are processed to eliminate the noise from the signals added during the acquisition using certain techniques such as Notch and band-pass filters, sample reduction, and Independent Component Analysis (ICA). 

After that, features are extracted from the signals, allowing ML algorithms to classify the\addtxt{se} \change{features}{characteristics} according to patterns identified in the data and, therefore, to predict drowsiness. \addtxt{It is relevant to highlight that Deep Learning (DL) is gaining popularity in identifying drowsiness while driving. However, DL approaches present several disadvantages, such as the amount of data required to train the models, the limited speed in training and evaluating models compared to traditional ML approaches, or the difficulty in explaining the decisions of the model \cite{Ahmed:CNN-bulky:2021, Zhang:survey-DL:2021}.}

Despite the advances and contributions of existing studies combining BCIs and ML to detect drowsiness while driving, there is a lack of literature analyzing the performance of customized and heterogeneous ML algorithms. The current literature presents a substantial amount of studies using ML, but in most of them, Support Vector Machine (SVM) is used without analyzing and comparing other well-known and relevant algorithms. In addition, the state of the art only explores the performance of customized and individual models trained with data from single subjects, presenting significant scalability issues for new subjects since a new training process per user is needed. In this sense, scalable models combining the brain activity of several subjects should be explored and analyzed to determine if they effectively detect sleepiness in various subjects, even if the models were not trained with their data.

To improve the previous challenges, this work presents the following main contributions:

\begin{itemize}
\item The design of a BCI and ML-based framework for drowsiness detection in driving scenarios employing EEG and Electrooculography (EOG) as features. The proposed framework considers ML classifiers and regressors for detecting different drowsiness levels in both individual users and groups of them.

\item The creation of a personalized algorithm for Percentage of Eye Closure (PERCLOS) discretization to improve drowsiness labeling, which takes into account the subject behavior to establish the thresholds between three drowsiness levels. 

\item The deployment and evaluation of the framework using a publicly available dataset, SEED-VIG \cite{seed-vig-dataset}, modeling the EEG of 21 subjects while driving. The following ML algorithms have been trained and evaluated with different amounts of subjects and features for regression and three-class classification tasks: SVM, k-Nearest Neighbors (kNN), Decision Trees (DT), Random Forest (RF), and Gaussian Processes (GP).

\item The obtained results indicate that algorithms such as RF or kNN\change{offer better results than the most common one in the literature, namely SVM}{, which are not widely explored in the literature, can improve the performance of the most commonly used techniques, such as SVM}. In particular, within individual models, RF performed the best with a mean f1-score of 78\% compared to SVM with 58\%. Similarly, RF is also the \change{best}{most promising} alternative for scalable models, reaching an f1-score of 79\% while SVM \change{gets}{obtained} 52\%.
\end{itemize}

\addtxt{Despite the contributions of this work, it presents limitations in terms of the amount of data used to train ML models. Using richer datasets would be useful to generalize the results obtained to a greater portion of the population, allowing to explore more complex intelligent approaches, such as DL approaches.}

The rest of this paper is organized as follows. Section~\ref{sec:2_realted_work} presents the state of the art from drowsiness detection in driving scenarios using BCIs. Subsequently, Section~\ref{sec:3_proposed_solution} presents the design of the proposed framework, followed by Section~\ref{sec:4_results} which states the results of detecting drowsiness using the framework. Finally, Section~\ref{sec:5_conclusions} presents \change{the conclusion}{some conclusions} and potential future work. 

\section{Related Work}\label{sec:2_realted_work}
This section analyzes how drowsiness assessment techniques using BCIs are implemented in the literature and what methodology is followed by each study. In particular, it \change{analyzes}{documents} what \addtxt{biosignals and data processing data are utilized, what} features are extracted from the \change{subjects}{signals}, the algorithms and models used to classify the signals, and \change{how well they perform}{their performance}. In the literature, both drowsiness and fatigue are related to the same concept of a person’s tendency to fall asleep. Every study analyzed shares the same starting point, an existing dataset. Some of them decide to generate their own \change{one}{data}, \change{whereas}{while} others opt for a public dataset \cite{Akbar:Recurrent-SVM:2019,Gwak:SVM-kNN-RF:2018,Lin:SVM-kernel-survey:2014}. \addtxt{After that, it is necessary to apply data processing techniques to improve the quality of the signals, such as removing artifacts \cite{Zhu:PSD-CNN:2021, Cui:Oz-survey:2021}.}

\change{Moreover, f}{F}eatures are \addtxt{then} extracted from \change{three}{different} sources\change{, each one corresponding}{. In the case of EEG, each source corresponds} to a transformed domain where EEG signals can be studied. Each study analyzed chooses certain features that may differ from the rest. Firstly, time-domain features are based on mathematical models and other algebraic operations, where the most popular and widespread is the Autoregressive Model (AR) \cite{Garces:ANN:2014, Chakladar:TD-SVM-RF:2020}. It is also common to extract statistical values from the signals such as variance, standard deviation and quantiles, or Hjorth parameters \cite{Akbar:Recurrent-SVM:2019,Chakladar:TD-SVM-RF:2020,Cui:Oz-survey:2021}.

aThe second source \addtxt{commonly used in EEG research} comes from frequency-domain features, where Fast Fourier Transform (FFT) enables the analysis of the predominant frequencies in the original EEG signals and their amplitude. Using FFT, the Power Spectral Density (PSD) is widely employed to measure the energy in each frequency band of the brain signals, providing good results when estimating drowsiness \cite{Akbar:Recurrent-SVM:2019,Wei:PSD-survey:2018,Shen:TN:2021,Zhu:PSD-CNN:2021,Cheng:SVM-CNN:2018}.

Thirdly, time-frequency domain features, due to the non-stationary, non-linear and non-Gaussian behavior exhibited by the EEG signals, are useful to have a representation and decomposition of the frequency information of the signals linked to the \change{time}{temporal} domain. This is why methods such as Discrete Wavelet Transform (DWT) are used \cite{Garces:ANN:2014,Chen:DWT-SVM-ELM:2015}. In addition to EEG features, it is common to combine them with other features which are extracted from the subject's behavior. These include heart rate (HR), blink rate \rmvtxt{(BR)}, or the number of blinks \cite{Naurois:adapted-ANN:2018,Hu:SVM:2009}. \addtxt{The blink rate determines the frequency or speed of blinking, while the number of blinks refers to the total number of blinks performed within a particular time interval.}

Finally, after feature extraction, the signals are classified. \change{There are two aspects in common while classifying in the analyzed studies}{There are two common aspects in the analyzed studies while classifying}. First, most works use a supervised learning approach and, second, they use a limited range of algorithms which are known to \change{give}{provide} good results, being SVM the most popular and widespread \change{one}{technique} \cite{Chen:DWT-SVM-ELM:2015,Mervyn:SVM:2009,Akbar:Recurrent-SVM:2019,Lin:SVM-kernel-survey:2014, Hu:SVM:2009}. This algorithm is followed in popularity by linear models, such as Ridge Regression, Logistic Regression, Lasso Regression, Naive Bayes and kNN \cite{Cui2019}. To a lesser extent, and with more popularity in other \change{branches}{areas} of EEG analysis, Linear Discriminant Analysis (LDA), DT and RF are also chosen \cite{Gwak:SVM-kNN-RF:2018,Wei:PSD-survey:2018}.

Regarding Deep Learning (DL), the most widely used neural networks are Convolutional Neural Networks (CNN), Long Short-Term Memory (LSTM), Extreme Learning Machines (ELMs) and Recurrent Self-Evolving Fuzzy Neural Networks (RSEFNNs). They are gaining relevance as they produce better results, in many cases, compared to traditional ML methods in drowsiness assessment \cite{Chen:DWT-SVM-ELM:2015, Naurois:adapted-ANN:2018,Garces:ANN:2014}.

When estimating sleepiness with supervised learning, the labels used for regression models are the values measured by \addtxt{self-assessment test, such as} KSS, NASA-TLX, \addtxt{and} Auditory Vigilance Task (AVT)\addtxt{,} or PERCLOS \addtxt{values}. If the problem is approached with a classification model, the values of the labels used in the regressive methods are discretized to different levels of drowsiness \cite{Zhuang:PERCLOS:2020,Savas:PERCLOS:2020,Chakladar:TD-SVM-RF:2020}.

\subsection{Performance of Literature Works}
This section presents an in-depth examination of the literature to identify how the algorithms perform while also \change{taking into account}{considering} the \addtxt{processing techniques and} features \rmvtxt{that} researchers adopt when estimating drowsiness. \addtxt{Focusing first on works employing ML approaches, Chen et al. \cite{Chen:DWT-SVM-ELM:2015} acquired EEG and EOG signals from 16 subjects using a nine-electrode BCI with a sampling rate of 256 Hz. Then, neurologists removed data artifacts and labeled the signals by visual inspection. Moreover, the authors extracted features from EEG using Discrete Wavelet Transformations (DWT) and combined them with EOG features. After that, the authors used SVM for classification, reaching an accuracy of 94.7\%.}

\addtxt{Gwak et al. \cite{Gwak:SVM-kNN-RF:2018} used ML to detect drowsiness at the wheel, analyzing different physiological signals and driving behaviors in a driving simulation for 16 subjects. This work used a 16-channel BCI with a sampling rate of 500 Hz, applying a band-pass filter between 1-40 Hz and ICA to remove artifacts. The authors considered 32 features obtained from PSD in EEG signals, ECG characteristics, eye movement, seat pressure, and driving simulation parameters. This study trained LR, SVM, kNN, and RF classifiers, where RF obtained 81.4\% accuracy in binary classification, in contrast to SVM, which obtained 78.6\% accuracy.} 

\addtxt{The work performed by Li et al. (2018) \cite{Li:SEED-VIG-SVR:2018} is relevant to the present study since the dataset employed is also SEED-VIG. The authors applied ICA and downsampling of 125 Hz to the EEG signals, obtaining 100 features related to differential entropy, while this work obtained 36 EOG features from horizontal and vertical channels. After that, the paper employed a Support Vector Regressor (SVR) as a baseline, resulting in a model with an RMSE of 0.17 and CC of 0.76.}
    
\addtxt{Wei et al. \cite{Wei:PSD-survey:2018} used a 32-channel BCI with a sampling rate of 500 Hz to acquire EEG signals from ten participants, utilized to predict drowsiness in a virtual driving environment. The authors processed the EEG using a band-pass filter between 1-50 Hz, a notch filter at 60 Hz, downsampling to 250 Hz, and Artifact Subspace Reconstruction (ASR). This work employed three-second epochs to obtain PSD features from theta, alpha, and beta waves from EEG. This research used LDA, kNN, and SVM algorithms for classification, where SVM obtained the best results, with an accuracy of 80\%.}

\addtxt{Akbar and Igasaki \cite{Akbar:Recurrent-SVM:2019} used an EEG BCI with 19 electrodes and a sampling rate of 500 Hz. The authors applied a band-pass filter between 0.5–50 Hz, then extracted Hjorth parameters and PSD from the frequency domain and KSS to self-assess drowsiness. The algorithm for classifying was SVM, achieving an RMSE of 0.15 and a $R^2$ of 0.83.} 

\addtxt{Qian et al. \cite{Qian:Drowsiness:2021} studied the detection of drowsiness during daytime short naps using EEG data obtained from 25 subjects with a sampling rate of 100 Hz. The authors selected frequencies under 30 Hz and then extracted features from EEG frequency bands using FFT. Finally, the authors studied several models, where the most promising alternative was a Bayesian-Copula Discriminant Classifier (BCDC) with 94.3\% accuracy, followed by Gaussian SVM (GSVM) with 93.7\% accuracy.}

\addtxt{Arefnezhad et al. \cite{Arefnezhad:Encoding:2022} proposed an encoder-decoder method for drowsiness detection in driving scenarios, using EEG signals obtained from 13 subjects using a BCI with eight channels and a sampling rate of 500 Hz. This work used ICA for data processing, then extracting EEG features consistent between subjects: skewness of alpha, delta power, theta power, and Hjorth mobility of delta. For classification, the encoder uses a series of equations to relate the EEG features obtained with PERCLOS values, resulting in relevant biomarkers in the EEG. In contrast, the decoder uses Bayes filtering and biomarkers to predict PERCLOS values.}

\addtxt{Arif et al. \cite{Arif:ML:2023} utilized various ML algorithms to detect drowsiness. In particular, they used a BCI device with 16 channels and a sampling rate of 125 Hz on 12 subjects. Then, this work used a band-pass filter between 0.5-40 Hz and a notch filter on 50 and 60 Hz frequencies, obtaining eight features from PSD and four from the band power ratio indices. Finally, they used decision trees, discriminant analysis, logistic regression, Naïve Bayes, SVM, kNN, and an ensemble classifier (bagged trees) for classification. This work concluded that the best approach was using an ensemble classifier, obtaining 85.6\% accuracy.}

\addtxt{Besides traditional ML approaches, the literature has explored the use of DL. In particular, Chakladar et al. \cite{Chakladar:TD-SVM-RF:2020} performed a workload analysis, exploring both ML and DL approaches. This work used an EEG dataset with 14 channels, with a sampling rate of 128 Hz, obtained from 48 participants. The subjects were recorded when no task was performed and during a simultaneous capacity multitasking activity, identifying three workload levels: low, moderate, and high. This research applied a band-pass of 4-32 Hz over the EEG, then extracted different features: PSD, hurst exponent, signal statistics (mean, standard deviation, skewness, kurtosis), approximate entropy, and autoregressive coefficient. Finally, three classification algorithms were tested: SVM, RF, and a hybrid approach of a Long Short-Term Memory (LSTM) with a Bidirectional LSTM, known as BLSTM-LSTM. This latter algorithm offered the best results, with 86.33\% accuracy. Moreover, Cheng et al. \cite{Cheng:SVM-CNN:2018} compared the performance of SVM with a CNN using EEG signals. The former obtained an accuracy of 64.05\%, while the latter achieved an accuracy of 69.19\%. In both cases, PSD features were utilized.}

\addtxt{Cui et al. \cite{Cui:Oz-survey:2021} used an explainable CNN with data from 27 subjects to detect drowsiness. The BCI has 32 electrodes and a sampling rate of 500 Hz. After the acquisition, the data is band-filtered between 1-50 Hz, removing artifacts using AAR, following a downsampling process to 128 Hz. After that, three-second epochs are used as raw inputs to a CNN network. The results, calculated for each subject, present an overall inter-subject accuracy of 73.22\%.}

\addtxt{Paulo et al. \cite{Paulo:DL:2021} used EEG signals from 27 subjects obtained from a 32-channel BCI with a sampling rate of 500 Hz. The authors applied a band-pass filter between 1-50 Hz and blink and muscular artifacts using AAR. This work trained a CNN with one convolutional layer and three dense layers with three-second temporal windows. The drowsiness problem was approached as an image classification problem, where the images represent spatiotemporal image encoding representations in the form of recurrence plots or gramian angular fields. The overall performance between individual models was 75.87\% accuracy.}

\addtxt{Shen et al. \cite{Shen:TN:2021} evaluated multiple DL approaches to detect drowsiness in driving scenarios. This study used data obtained from a 32-channel EEG BCI with a sampling rate of 500 Hz, obtained from 11 subjects. After performing a band-pass filter between 1-50 Hz, Automatic Artifact Removal (AAR), and downsampling to 120 Hz, the authors calculated the PSD over each EEG channel, labeling the different experimental sessions as drowsy or alert. This work explored different classification approaches, where the most promising was their proposed method, consisting in the use of multi-source signal alignment with a tensor network, reaching a 71.97\% accuracy in leave-one-subject-out cross-validation.} 

\addtxt{Zhu et al. \cite{Zhu:PSD-CNN:2021} used an eight-channel EEG-based BCI with a sampling rate of 256 Hz on ten subjects. This article applied a band-pass filter between 1-60 Hz, a notch filter on 50 Hz, ICA, and the wavelet threshold method. Then, the authors trained a CNN to predict drowsiness while driving, evaluating either the application of an Inception or an AlexNet module. The use of the Inception module offered an accuracy of 95.59\%, while the use of the AlexNet approach reached 94.68\% accuracy.}

\rmvtxt{In particular, it is worth highlighting the work performed by Chen et al. \cite{Chen:DWT-SVM-ELM:2015}, where neurologists removed data artifacts and labeled the signals by visual inspection. Moreover, the authors used SVM for classification, reaching an accuracy of 94.7\%, using DWT and eye features as well. In terms of regression, Akbar and Igasaki \cite{Akbar:Recurrent-SVM:2019} achieved 0.15 RMSE and 0.83 $R^2$. SVM was again the algorithm of choice. This publication used time-domain features such as Hjorth parameters and PSD from the frequency-domain, and KSS to self-assess drowsiness.}

\rmvtxt{The work done by Li et al. (2018) \cite{Li:SEED-VIG-SVR:2018} is relevant for the present study since the dataset employed is also SEED-VIG. These author employed a Support Vector Regressor (SVR) as baseline together with 100 EEG and 36 EOG features, resulting in a model that achieved an RMSE of 0.17$\pm$0.06 and CC of 0.76$\pm$0.23.}

\rmvtxt{Studies like \cite{Gwak:SVM-kNN-RF:2018, Cui:Oz-survey:2021, Wei:PSD-survey:2018} analyzed different algorithms, being some of them not so popular in the literature, such as kNN, DT or RF. In those studies, SVM is also included and, in some cases, there are other algorithms offering better results. Furthermore, in \cite{Gwak:SVM-kNN-RF:2018}, RF obtained 81.4\% while SVM only reached 78.6\%. These results highlight that, although SVM is the most widespread algorithm, there are alternatives which presented promising results in their studies.}

\rmvtxt{Regarding DL, its use is increasing in these studies and often produces results surpassing traditional ML algorithms. An example of this is shown by Cheng et al. \cite{Cheng:SVM-CNN:2018}, who obtained and accuracy of 69.19\% with CNN, compared to 64.05\% employing SVM, using the sames features and labeling in both approaches.}

In particular, \tablename~\ref{tab:related-work-results} presents a summary of the \change{examined}{most related} studies, specifying the \addtxt{processing techniques,} set of features, data labeling and the algorithms used. If a work uses regression, the results are expressed by the Root Mean Square Error (RMSE)\rmvtxt{, with results between 0.15 and 0.5}. In addition to RMSE, the Pearson correlation coefficient (CC) or the coefficient of determination ($R^2$) is used \addtxt{depending on the paper}. On the other hand, classification models are characterized by accuracy as performance metric. After analyzing the literature it can be seen that SVM is generally present in the literature but, at the same time, there is evidence of other approaches, such as neural networks or other ML algorithms, that offer similar or even better results. In addition, there is also a lack of studies that consider scalable models since all of the identified studies focus on individual models which can only detect drowsiness in a specific subject.

\begin{table*}[!htb]
\centering
\caption{Summary of the literature works reviewed and their results. \addtxt{The results are expressed in multiple metrics, where \textit{Acc} represents Accuracy, \textit{CC} means the Pearson correlation coefficient, \textit{RMSE} is the Root Mean Square Error, and \textit{$R^2$} indicates the coefficient of determination.}}
    \label{tab:related-work-results}
\resizebox{\textwidth}{!}{
\begin{tabular}{@{}llllll@{}}
\toprule
\textbf{Reference} & \textbf{Acquisition} & \textbf{Processing} & \textbf{Features} & \textbf{Classification} & \textbf{Results} \\ \midrule

\begin{tabular}{@{}c@{}}Chen et al. \\ (2015) \cite{Chen:DWT-SVM-ELM:2015}\end{tabular} & \addtxt{EEG, EOG} & \addtxt{N/A} & Four from DWT, blinking & \begin{tabular}{@{}l@{}}SVM \\ ELM \end{tabular} & \begin{tabular}{@{}l@{}}\addtxt{Acc: }96.90\% \\ \addtxt{Acc: }97.30\% \end{tabular} \\ \midrule

\begin{tabular}{@{}c@{}}Cheng et al. \\ (2018) \cite{Cheng:SVM-CNN:2018}\end{tabular} & \addtxt{EEG} & \addtxt{N/A} & PSD & \begin{tabular}{@{}l@{}}SVM \\ CNN \end{tabular} & \begin{tabular}{@{}l@{}}\addtxt{Acc: }64.05\% \\ \addtxt{Acc: }69.19\% \end{tabular} \\ \midrule

\begin{tabular}{@{}c@{}}Gwak et al. \\ (2018) \cite{Gwak:SVM-kNN-RF:2018}\end{tabular} & \addtxt{EEG, ECG, EOG} & \addtxt{Band-pass 1-40 Hz, ICA} & \begin{tabular}{@{}l@{}} PSD, ECG, EOG \\ simulation data \end{tabular} & \begin{tabular}{@{}l@{}}CNN \\ SVM \\ RF \end{tabular} & \begin{tabular}{@{}l@{}}\addtxt{Acc: }75.30\% \\ \addtxt{Acc: }78.60\% \\ \addtxt{Acc: }81.40\% \end{tabular} \\ \midrule

\begin{tabular}{@{}c@{}}Li et al. \\ (2018) \cite{Li:SEED-VIG-SVR:2018}\end{tabular} & \addtxt{EEG, EOG} & \addtxt{ICA, downsampling 125 Hz} & \begin{tabular}{@{}l@{}} 100 from EEG, \\ 36 from EOG \end{tabular} & \begin{tabular}{@{}l@{}}SVR \end{tabular} & \begin{tabular}{@{}l@{}}CC: 0.76 \\ RMSE: 0.17 \end{tabular} \\ \midrule

\begin{tabular}{@{}c@{}}Wei et al. \\ (2018) \cite{Wei:PSD-survey:2018}\end{tabular} & \addtxt{EEG} & \begin{tabular}{@{}c@{}}\addtxt{Band-pass 1-50 Hz, notch} \\ \addtxt{60 Hz, downsampling 250} \\ \addtxt{Hz, ASR}\end{tabular} & \begin{tabular}{@{}l@{}} PSD \end{tabular} & \begin{tabular}{@{}l@{}}kNN \\ LDA \\ SVM\end{tabular} & \begin{tabular}{@{}l@{}}\addtxt{Acc: }77.3\% \\ \addtxt{Acc: }79.4\% \\ \addtxt{Acc: }80.0\%\end{tabular} \\ \midrule

\begin{tabular}{@{}c@{}}Akbar and \\ Igasaki \\ (2019) \cite{Akbar:Recurrent-SVM:2019}\end{tabular} & \addtxt{EEG} & \addtxt{Band-pass 0.5-50 Hz} & \begin{tabular}{@{}l@{}} Hjorth parameters, PSD \end{tabular} & \begin{tabular}{@{}l@{}}SVM \\ Recurrent SVM \end{tabular} & \begin{tabular}{@{}l@{}} $R^2$: 0.64, RMSE: 0.56 \\ $R^2$: 0.83, RMSE: 0.15 \end{tabular} \\ \midrule

\begin{tabular}{@{}c@{}}Chakladar et \\ al. (2020) \cite{Chakladar:TD-SVM-RF:2020}\end{tabular} & \addtxt{EEG} & \addtxt{Band-pass 5-32 Hz} & \begin{tabular}{@{}c@{}}PSD, mean, SD, skewness, \\ kurtosis, AR, entropy\end{tabular} & \begin{tabular}{@{}l@{}}RF \\ SVM\end{tabular} & \begin{tabular}{@{}l@{}}\addtxt{Acc: }83.00\% \\ \addtxt{Acc: }83.33\% \end{tabular} \\ \midrule

\begin{tabular}{@{}c@{}}Cui et al. \\ (2021) \cite{Cui:Oz-survey:2021}\end{tabular} & \addtxt{EEG} & \begin{tabular}{@{}c@{}}\addtxt{Band-pass 1-50 Hz, AAR,} \\ \addtxt{downsampling 128 Hz}\end{tabular} & \begin{tabular}{@{}c@{}}Oz channel\end{tabular} & \begin{tabular}{@{}l@{}}DT \\ RF \\ kNN \\ Gaussian Naïve Bayes \\ SVM \end{tabular} & \begin{tabular}{@{}l@{}} \addtxt{Acc:} 60.70\% \\ \addtxt{Acc: }62.60\% \\ \addtxt{Acc: }63.42\% \\ \addtxt{Acc: }67.44\% \\ \addtxt{Acc: }69.72\% \end{tabular} \\ \midrule

\begin{tabular}{@{}c@{}}\addtxt{Paulo et al.} \\ \addtxt{(2021) \cite{Paulo:DL:2021}}\end{tabular} & \addtxt{EEG} & \begin{tabular}{@{}l@{}}\addtxt{Band-pass 1-50 Hz, AAR}\end{tabular} & \addtxt{Raw EEG} & \begin{tabular}{@{}l@{}}\addtxt{CNN} \end{tabular} & \begin{tabular}{@{}l@{}}\addtxt{Acc: 75.87\%} \end{tabular} \\ \midrule

\begin{tabular}{@{}c@{}}Shen et al. \\ (2021) \cite{Shen:TN:2021}\end{tabular} & \addtxt{EEG} & \begin{tabular}{@{}c@{}}\addtxt{Band-pass 1-50 Hz, AAR,} \\ \addtxt{downsampling 120 Hz}\end{tabular} & PSD & \begin{tabular}{@{}l@{}}SVM \end{tabular} & \begin{tabular}{@{}l@{}}\addtxt{Acc: }62.51\% \end{tabular} \\ \midrule

\begin{tabular}{@{}c@{}}\addtxt{Qian et al.} \\ \addtxt{(2021) \cite{Qian:Drowsiness:2021}}\end{tabular} & \addtxt{EEG} & \begin{tabular}{@{}c@{}}\addtxt{Band-pass $<$30 Hz}\end{tabular} & \addtxt{PSD} & \begin{tabular}{@{}l@{}}\addtxt{BCDC} \\ \addtxt{GSVM} \end{tabular} & \begin{tabular}{@{}l@{}}\addtxt{Acc: 94.3\%} \\ \addtxt{Acc: 93.7\%} \end{tabular} \\ \midrule

\begin{tabular}{@{}c@{}}Zhu et al. \\ (2021) \cite{Zhu:PSD-CNN:2021}\end{tabular} & \addtxt{EEG} & \begin{tabular}{@{}c@{}}\addtxt{Band-pass 1-60 Hz, notch} \\ \addtxt{50 Hz, ICA, wavelet threshold}\end{tabular} & Raw EEG & \begin{tabular}{@{}l@{}}CNN-Inception \\ CNN-AlexNet \end{tabular} & \begin{tabular}{@{}l@{}}\addtxt{Acc: }93.6\% \\ \addtxt{Acc: }94.68\% \end{tabular} \\ \midrule

\begin{tabular}{@{}c@{}}\addtxt{Arefnezhad et} \\ \addtxt{al. (2022)} \cite{Arefnezhad:Encoding:2022}\end{tabular} & \begin{tabular}{@{}c@{}}\addtxt{EEG}\end{tabular} & \addtxt{ICA} & \begin{tabular}{@{}c@{}}\addtxt{Skewness alpha, delta power,} \\ \addtxt{theta power, Hjorth delta mobility} \end{tabular} & \begin{tabular}{@{}l@{}}\addtxt{Encoder-decoder} \end{tabular} & \begin{tabular}{@{}l@{}}\addtxt{RMSE: 0.117} \end{tabular} \\ \midrule

\begin{tabular}{@{}c@{}}\addtxt{Arif et al.} \\ \addtxt{(2023)} \cite{Arif:ML:2023}\end{tabular} & \begin{tabular}{@{}c@{}}\addtxt{EEG}\end{tabular} & \begin{tabular}{@{}c@{}}\addtxt{Band-pass} \\ \addtxt{0.5-40Hz, Notch} \end{tabular} & \begin{tabular}{@{}c@{}}\addtxt{PSD, band power ratio indices} \end{tabular} & \begin{tabular}{@{}l@{}} \addtxt{Discriminant analysis} \\ \addtxt{LR} \\ \addtxt{Naïve Bayes} \\ \addtxt{SVM} \\ \addtxt{DT} \\ \addtxt{kNN} \\ \addtxt{Ensemble classier} \end{tabular} & \begin{tabular}{@{}l@{}}\addtxt{Acc: 63.5\%} \\ \addtxt{Acc: 63.6\%} \\ \addtxt{Acc: 67.4\%} \\ \addtxt{Acc: 75.7\%} \\ \addtxt{Acc: 77.4\%} \\ \addtxt{Acc: 78.5\%} \\ \addtxt{Acc: 85.6\%}\end{tabular} \\

\bottomrule
\end{tabular}}
\end{table*}

\section{Proposed Solution}\label{sec:3_proposed_solution}
\begin{figure}[ht]
\centering
    \includegraphics[width=\columnwidth]{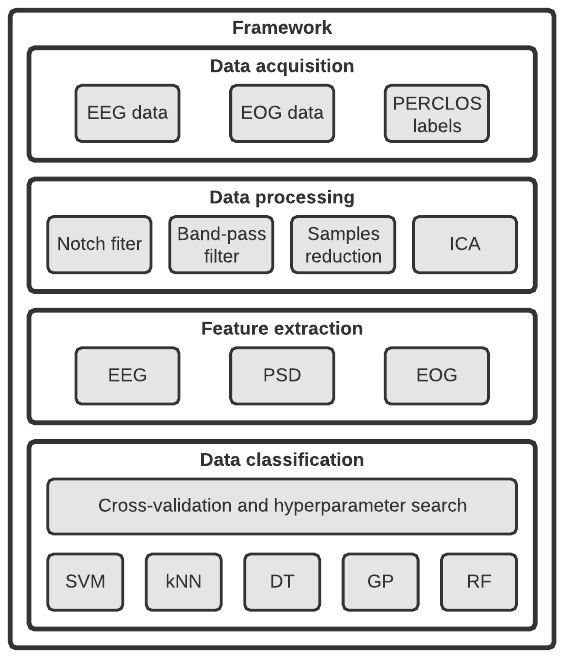}
    \caption{Framework overview.}
    \label{fig:framework-overview}
\end{figure}

This section describes the design and implementation of the proposed framework to detect drowsiness while driving\addtxt{, related to the first contribution indicated in the Introduction section}. An overview of the framework is shown in \figurename~\ref{fig:framework-overview}, presenting its different components. Starting from the upper side, the first two components refer to the acquisition of data and its processing. Next, a feature extraction stage selects the most relevant aspects of the acquired data. Finally, the framework includes a data classification block, where individual models for each subject and scalable models with data from several users are implemented based on different ML algorithms.

\addtxt{This framework differentiates from existing platforms, focusing on the particularities of EEG and EOG signals offering specific processing capabilities for drowsiness detection. Moreover, the framework implements a novel PERCLOS discretization approach able to adapt to the particularities of each subject. Finally, the proposed framework tests a substantial variety of ML algorithms to offer a detailed comparison between them in terms of well-known performance metrics.}

\addtxt{It is worth mentioning that the structure of the proposed framework is aligned with existing frameworks using EEG signals to predict particular dynamics of the human brain \cite{Martinez-Beltran:framework:2022, Xu:Framework:2019}. Moreover, the modules of these frameworks have a direct association with the phases of the BCI life cycle and traditional ML methodologies, which represent the stages required to acquire biosignals from the brain, their transformation to be understood by computers, and, finally, the use of learning techniques to predict specific events within the signals \cite{Lopez_Bernal:cyberBCI:2021}.}


\subsection{Data Acquisition}
The design and implementation of the proposed framework is generic enough to be compatible with different datasets, as well as data coming in real-time from a BCI. Nevertheless, this work used the SEED-VIG dataset \cite{seed-vig-dataset} due to its realistic conditions, the suitability with the study purpose, and the amount and quality of the data provided.

\begin{figure}[ht]
\centering
    \includegraphics[width=0.85\columnwidth]{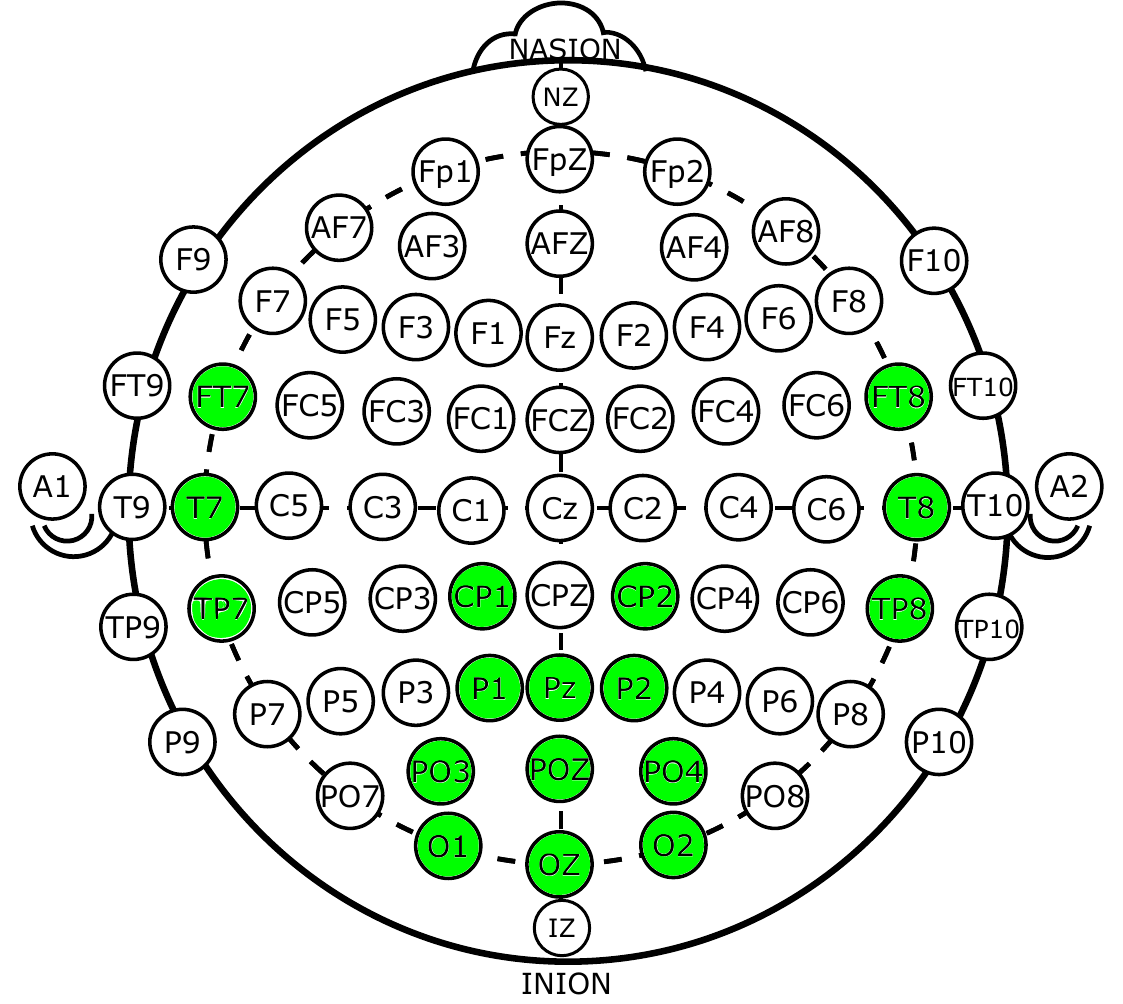}
    \caption{Placement of the EEG electrodes used in the SEED-VIG dataset, highlighted in color.}
    \label{fig:electrodes-placement}
\end{figure}

More in detail, the SEED-VIG dataset consists of 23 experiments over 21 different subjects (two subjects repeated the experiment). Each experiment has about two hours of EEG signals recorded while the subjects were using a driving simulator. The experiments acquired data from 17 electrode channels according to the 10-20 system (see \figurename~\ref{fig:electrodes-placement}), using a sample rate of 200 Hz. Particularly, the \textit{Neuroscan} BCI device was in charge of acquiring EEG and EOG biosignals \cite{bci-dataset}. This dataset provides the raw data from the different experiments, together with a variety of already processed data. Particularly, the present study uses the following data subsets: 1) raw EEG data from the 17 EEG channels, 2) average PSD relative to the five frequency bands of the brain signals and, 3) raw data from the EOG vertical channel.

The dataset was labeled every eight seconds with subjects' PERCLOS values obtained by an eye-tracking device from \textit{SensoMotoric Instruments} \cite{glasses-dataset}. PERCLOS is a psycho-physiological measure of the subject that quantifies the percentage of time that a subject has been with the eyes at least 80\% closed during the time interval of measurement \cite{perclos-definition}.

\subsection{Data Processing}
\begin{figure}[ht]
\centering
    \includegraphics[width=0.85\columnwidth]{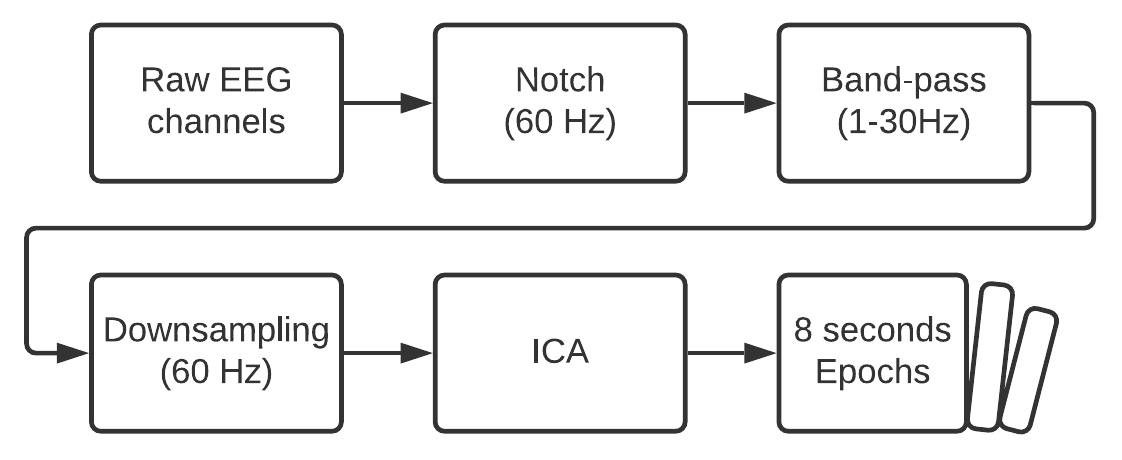}
    \caption{EEG signals processing phase.}
    \label{fig:EEG-filtering}
\end{figure}

As a result of using a non-invasive BCI, the EEG signals obtained contain artifacts, so they must be filtered following the process presented in \figurename~\ref{fig:EEG-filtering}. Initially, the signals are processed with two filtering techniques. First, a Notch filter applied at 60 Hz eliminates the noise introduced by the power grid. Secondly, a band-pass filter between one and 30 Hz was applied since this is the frequency range of interest for the study of drowsiness \cite{Ward:book:2019}. The signals are then downsampled to 60 Hz following the Nyquist–Shannon sampling theorem to reduce the size of the data and speed up its subsequent classification without losing information. Finally, ICA permits to remove the remaining artifacts, such as subjects' eye blinks, while the essential information for detecting drowsiness is preserved. Once the artifacts are removed from the initial raw data, it is also necessary to split the signals in portions (Epochs) of eight seconds. This allows to perform a correct feature extraction since there is a PERCLOS value every eight seconds.

\subsection{Feature Extraction}
\tablename~\ref{tab:features-description} shows the sources from which features are extracted, their description, and the total number of features calculated. Focusing on EEG features, the eight extracted features for each channel are: mean, standard deviation, variance, 5th percentile, first quartile, median, third quartile, and 95th percentile. Thus, a total of $8\times17=136$ features are obtained. Moreover, five features using PSD are calculated, one per frequency band among the 17 EEG channels. Finally, this phase calculates a final EOG feature. Then, the classification stage of the framework receives three feature vectors corresponding to the different combinations of features tested: \change{the first,}{1)} the use of the 136 EEG features; \change{the second,}{2)} the use of the five PSD features alone; and \change{the third,}{3)} a combination of PSD and EOG features. 
\begin{table}[]
\centering

\caption{Feature sources and description for each Epoch.}
\label{tab:features-description}

\begin{adjustbox}{width=\columnwidth}
\begin{tabular}{L{1.5cm}L{7cm}}
    \hline
    \textbf{Feature sources} & \textbf{Description} \\
    \hline
    EEG & Eight features representing the behavior of the signal for each particular electrode in a summarized version via statistical measures. Therefore, 136 features are obtained. \\
    \hline
    PSD & Five features, each one corresponding to the averaged power of a particular frequency band in the 17 EEG channels.\\
    \hline
    EOG & One feature corresponding to the number of blinks from the subject.\\
    \hline
\end{tabular}
\end{adjustbox}
\end{table}

\subsection{PERCLOS Discretization Algorithm and Drowsiness Classification}
There are two main categories of supervised learning techniques: regression, which predicts numerical values (PERCLOS values in this study); and classification, which produces class assignments. Both categories are used in the framework since either approaches are used in the literature, thus facilitating subsequent comparison of the results.

Since PERCLOS values range from zero to one, it is necessary to map them into three levels of sleepiness, as recommended by Trejo et al. (2007) \cite{three-class} and Chang et al. (2007) \cite{psd-bands-AND-three-class}. Regarding the literature, fixed thresholds are commonly chosen to divide the PERCLOS range of values into the levels of sleepiness. Nevertheless, Gu et al. (2018) \cite{Gu:perclos-thresholds} stated that it is not possible to directly use the thresholds of other studies since they are related to the different detection methods used by different researchers, concluding that the PERCLOS thresholds should be obtained from experiments themselves. 

Based on the above, the proposed framework applies a dynamic PERCLOS discretization algorithm to calculate the thresholds between classes for each subject. With this algorithm, the physiological particularities of each subject are taken into account, thus obtaining a personalized division of drowsiness levels that improves data labeling. The threshold between the \textit{minor} and \textit{moderate} drowsiness levels (th\_minor) is calculated with Equation~(\ref{eq:moderate-threshold}) while the threshold between moderate and severe drowsiness levels (th\_moder) is obtained by Equation~(\ref{eq:severe-threshold}). \rmvtxt{In each equation, there is a fixed value obtained after testing different combinations to find the best option that divides the data for the subjects in the experiments.} 

\begin{dmath}\label{eq:moderate-threshold}
\texttt{th_minor}=min(PERCLOS) +\\ (max(PERCLOS)-min(PERCLOS))*0.125
\end{dmath}

\begin{dmath}\label{eq:severe-threshold}
    \texttt{th\_moder}=min(PERCLOS) +\\ (max(PERCLOS)-min(PERCLOS))*0.30
\end{dmath}

\addtxt{Concerning the static threshold values in the equations, the literature establishes values between 7.5\% and 15\% for the minor threshold \cite{Nguyen:Threshold:2015, Celecia:Threshold:2020}. In particular, this manuscript considers the work performed by Bowman et al. \cite{Bowman:Threshold:2008}, which defined a 12.5\% threshold as being an intermediate value within the range. For the moderate range, the literature documents values between 15\% and 30\% \cite{Nguyen:Threshold:2015, Celecia:Threshold:2020, Selvakumar:Threshold:2016}. Based on that, this publication opted for a conservative approach, selecting 30\% for Equation~\ref{eq:severe-threshold}.} \addtxt{These aspects are aligned with the second contribution of the paper, focused on the creation of a personalized PERCLOS algorithm.}

\begin{figure}[ht]
\centering
    \includegraphics[width=\columnwidth]{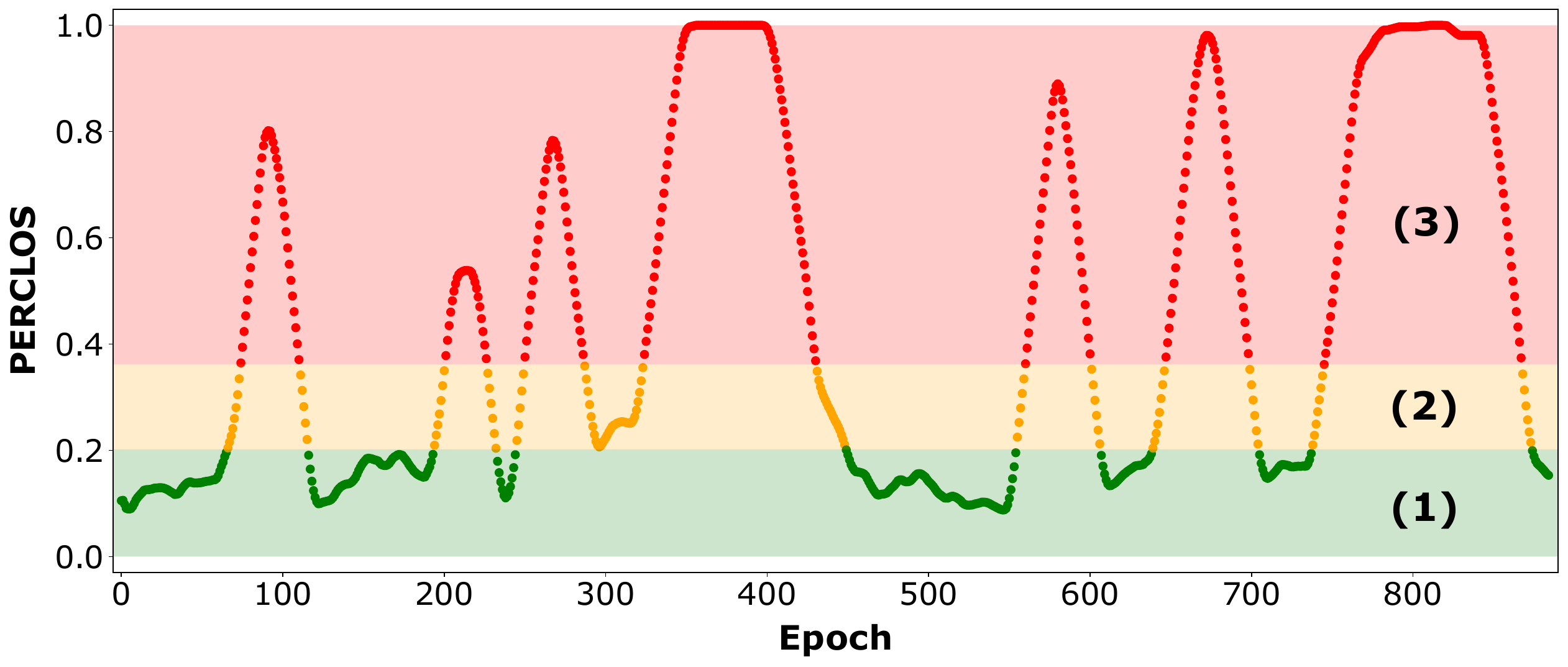}
    \caption{Output of the PERCLOS discretization algorithm with tree levels of drowsiness.}
    \label{fig:PERCLOS-discretization}
\end{figure}

A visual example of the PERCLOS discretization with the proposed algorithm in this study is shown in \figurename~\ref{fig:PERCLOS-discretization}. The green zone, marked as (1), contains the values where the subject's drowsiness is considered \textit{minor} or fully awake. Subsequently, the yellow zone, marked as (2), indicates \textit{moderate} drowsiness while the red zone, highlighted by (3), represents severe drowsiness.

During the classification stage, the framework uses two different \addtxt{ML} model approaches. The first one focuses on training individual and customized models for each user. The second category is based on training scalable models suitable for groups of subjects. Particularly, the two best-performing combinations in the individual models, together with the best one from SVM, are used for the scalable analysis, aiming to reduce the complexity of the experimentation. A combination is defined as a ML algorithm together with a vector of features. \addtxt{It is also essential to highlight that this framework does not implement DL algorithms due to the limitations indicated in the Introduction section, as the dataset used has a limited size.}

To train each model, the PSD and EOG features are normalized using a \textit{MinMax scaler}. Then, the framework shuffle\addtxt{s} the data before performing the splitting process, which varies according to the categories of models used. In individual models, the 75\% of the data defines the training set, while the remaining 25\% is used for testing. In contrast, the different combinations of scalable models have their own evaluation sample proportion. Moreover, ten-fold cross validation together with hyperparameter search allows finding the best configuration parameters of a model and achieving the best performance while avoiding overfitting. The algorithms of choice are SVM, kNN, DT, GP, and RF. From them, DT, kNN, and RF were selected based on the literature review previously presented, as these algorithms offer promising results. Finally, GP is selected because, although its behavior is non-Gaussian in contrast to EEG signals, it is interesting to evaluate its performance. \addtxt{It is worth mentioning that these methodological considerations are related to the third contribution presented in the Introduction.}

\section{Experiments and Results}\label{sec:4_results}

This section presents a set of experiments aiming to evaluate the drowsiness detection performance of individual and scalable ML models using regression and three-class classification techniques\addtxt{, covering the last contribution of the article}. Concerning trained models, there is one type of individual models while three types are explored for scalable models:
\begin{itemize}
    \item Individual models: \change{Individual}{Personalized} classifier and regressor trained and evaluated for each subject.
    
    \item 100 models: General classifier and regressor trained and evaluated with the 100\% of subjects\addtxt{, where the 75\% of the data across subjects is used for training and a 25\% for testing}.
    
    \item 90-10 models: General classifier and regressor trained with the 90\% of subjects and evaluated with the remaining 10\%.
    
    \item 70-30 models: General classifier and regressor trained with the 70\% of subjects and evaluated with the remaining 30\%.
\end{itemize}

Regarding regressive models, two metrics are used to measure the quality of the results: RMSE and $R^2$. \change{On the contrary}{Moreover}, four metrics allow to measure the performance of classification models: accuracy, precision, recall, and f1-score. Particularly, f1-score is prioritized because it involves both precision and recall, making it the most robust and meaningful metric for the analysis.

Since the results of individual, 90-10, and 70-30 models present multiple combinations of different algorithms and subjects tested, the results are presented averaged, subsequently indicated with the following format: \texttt{Mean}~$\pm$~\texttt{STD}. In contrast, there is no need to do this for 100 model\addtxt{s}, since there is only one test set with the data reserved from every experiment.

\subsection{Individual Models}
The performance of the trained regressive individual models is shown in \tablename~\ref{tab:individual-regression}, where the three assembled feature vectors (EEG, PSD, and EOG+PSD) are used to train each model to observe the performance of each one together with the different ML algorithms evaluated. Generally, it is observed that the lowest RMSE occurs in most cases when only the PSD features are used, followed closely by those using PSD together with EOG and, finally, those \change{using}{utilizing} only EEG data. It should also be noted that, although \rmvtxt{the} EEG \change{gives}{provides} the worst results in all cases, these results are acceptable to obtain a good prediction of sleepiness.

As expected, GP performs the worst for all three feature sets since this algorithm is based on the probabilistic theory of the Gaussian distribution, as discussed above. \change{On the other hand}{In contrast}, SVM and DT offer similar results in terms of their error, improving the results of GP. Finally, kNN and RF are the algorithms with the lowest RMSE. The combination offering the best performance is RF with PSD and EOG features both in RMSE ($0.08\pm0.02$) and $R^2$ ($0.83\pm0.09$).

\begin{table}[ht]
\centering

\caption{Regression performance for the individual models.}
\label{tab:individual-regression}

\resizebox{\columnwidth}{!}{\begin{tabular}{lcC{2.5cm}cc} 
    \hline
    \textbf{Algorithm} & \textbf{Features} & \textbf{RMSE} & \textbf{\boldmath$R^2$} \\ 
    \hline
    \multirow{3}{2cm}{SVM} & EEG & $0.16\pm0.05$ & $0.42\pm0.22$ \\
     & PSD & $0.12\pm0.04$ & $0.67\pm0.20$ \\
     & PSD+EOG & $0.12\pm0.04$ & $0.65\pm0.21$ \\ 
    \hline
    \multirow{3}{2cm}{kNN} & EEG & $0.15\pm0.05 $ & $0.49\pm0.20 $ \\
     & PSD & $0.09\pm0.03 $ & $0.82\pm0.09 $ \\
     & PSD+EOG & $0.10\pm0.04 $ & $0.75\pm0.18 $ \\ 
    \hline
    \multirow{3}{2cm}{DT} & EEG & $0.21\pm0.06$ & $0.06\pm0.39$ \\
     & PSD & $0.12\pm0.04$ & $0.68\pm0.19$ \\
     & PSD+EOG & $0.12\pm0.03$ & $0.68\pm0.17 $ \\ 
    \hline
    \multirow{3}{2cm}{GP} & EEG & $0.21\pm0.06 $ & $-0.07\pm0.69$ \\
     & PSD & $0.13\pm0.04 $ & $0.55\pm0.26 $ \\
     & PSD+EOG & $0.17\pm0.05 $ & $0.26\pm0.54 $ \\ 
    \hline
    \multirow{3}{2cm}{RF} & EEG & $0.14\pm0.05 $ & $0.56\pm0.17 $ \\
     & PSD & $0.09\pm0.03 $ & $0.83\pm0.09 $ \\
     &\textbf{PSD+EOG} & \boldmath$0.08\pm0.02 $ & \boldmath$0.83\pm0.09 $ \\ 
    \hline
\end{tabular}}
\end{table}

In the same way, the best combinations for classification are quite similar to those previously shown for regression since the algorithms are the same but focused on classification. Particularly, they are presented in \tablename~\ref{tab:individual-classification}. Nevertheless, \rmvtxt{it is worth commenting on the best combinations of the classification approach since} the metrics used are different and introduce a series of considerations that cannot be studied from the regressive point of view. \change{The three best combinations, regarding now f1-score, are again the same but in a different order. In this case, the best one is RF with PSD as features with an f1-score of $0.78\pm0.07$.}{In this case, RF with PSD obtains the best performance, with an f1-score of $0.78\pm0.07$, closely followed by kNN using PSD ($0.85\pm0.05$)}.

\begin{table}[ht]
\centering

\caption{Classification performance for the individual models.}
\label{tab:individual-classification}

\resizebox{\columnwidth}{!}{
\begin{tabular}{lcC{2.cm}cccc} 
    \hline
    \textbf{Algorithm} & \textbf{Features} & \textbf{Accuracy} & \textbf{Precision} & \textbf{Recall} & \textbf{f1-score} \\ 
    \hline
    \multirow{3}{1.8cm}{SVM} & EEG & $0.72\pm0.10$ & $0.71\pm0.16$ & $0.51\pm0.12$ & $0.50\pm0.13$ \\
     & PSD & $0.76\pm0.10$ & $0.78\pm0.12$ & $0.58\pm0.13$ & $0.58\pm0.14$ \\
     & PSD+EOG & $0.76\pm0.09$ & $0.074\pm0.12$ & $0.58\pm0.12$ & $0.58\pm0.13$ \\ 
    \hline
    \multirow{3}{1.8cm}{kNN} & EEG & $0.71\pm0.09$ & $0.61\pm0.13$ & $0.52\pm0.10$ & $0.51\pm0.11$ \\
     & PSD & $0.85\pm0.05$ & $0.81\pm0.50$ & $0.77\pm0.08$ & $0.78\pm0.07$ \\
     & PSD+EOG & $0.79\pm0.08$ & $0.73\pm0.12$ & $0.63\pm0.12$ & $0.64\pm0.12$ \\ 
    \hline
    \multirow{3}{1.8cm}{DT} & EEG & $0.67\pm0.11$ & $0.63\pm0.14$ & $0.47\pm0.09$ & $0.46\pm0.10$ \\
     & PSD & $0.80\pm0.08$ & $0.73\pm0.09$ & $0.71\pm0.08$ & $0.71\pm0.09$ \\
     & PSD+EOG & $0.80\pm0.09$ & $0.74\pm0.09$ & $0.72\pm0.07$ & $0.72\pm0.08$ \\ 
    \hline
    \multirow{3}{1.8cm}{GP} & EEG & $0.71\pm0.10$ & $0.70\pm0.13$ & $0.48\pm0.09$ & $0.46\pm0.09$ \\
     & PSD & $0.70\pm0.11$ & $0.79\pm0.17$ & $0.46\pm0.12$ & $0.42\pm0.13$ \\
     & PSD+EOG & $0.71\pm0.10$ & $0.77\pm0.12$ & $0.48\pm0.11$ & $0.45\pm0.12$ \\ 
    \hline
    \multirow{3}{1.8cm}{RF} & EEG & $0.74\pm0.09$ & $0.74\pm0.12$ & $0.54\pm0.10$ & $0.54\pm0.10$ \\
     & \textbf{PSD} & \boldmath$0.86\pm0.06$ & \boldmath$0.83\pm0.06$ & \boldmath$0.76\pm0.08$ & \boldmath$0.78\pm0.07$ \\
     & PSD+EOG & $0.85\pm0.06$ & $0.83\pm0.05$ & $0.74\pm0.08$ & $0.76\pm0.07$ \\ 
    \hline
\end{tabular}}
\end{table}

\begin{figure*}[ht]
\centering
    \includegraphics[width=\textwidth]{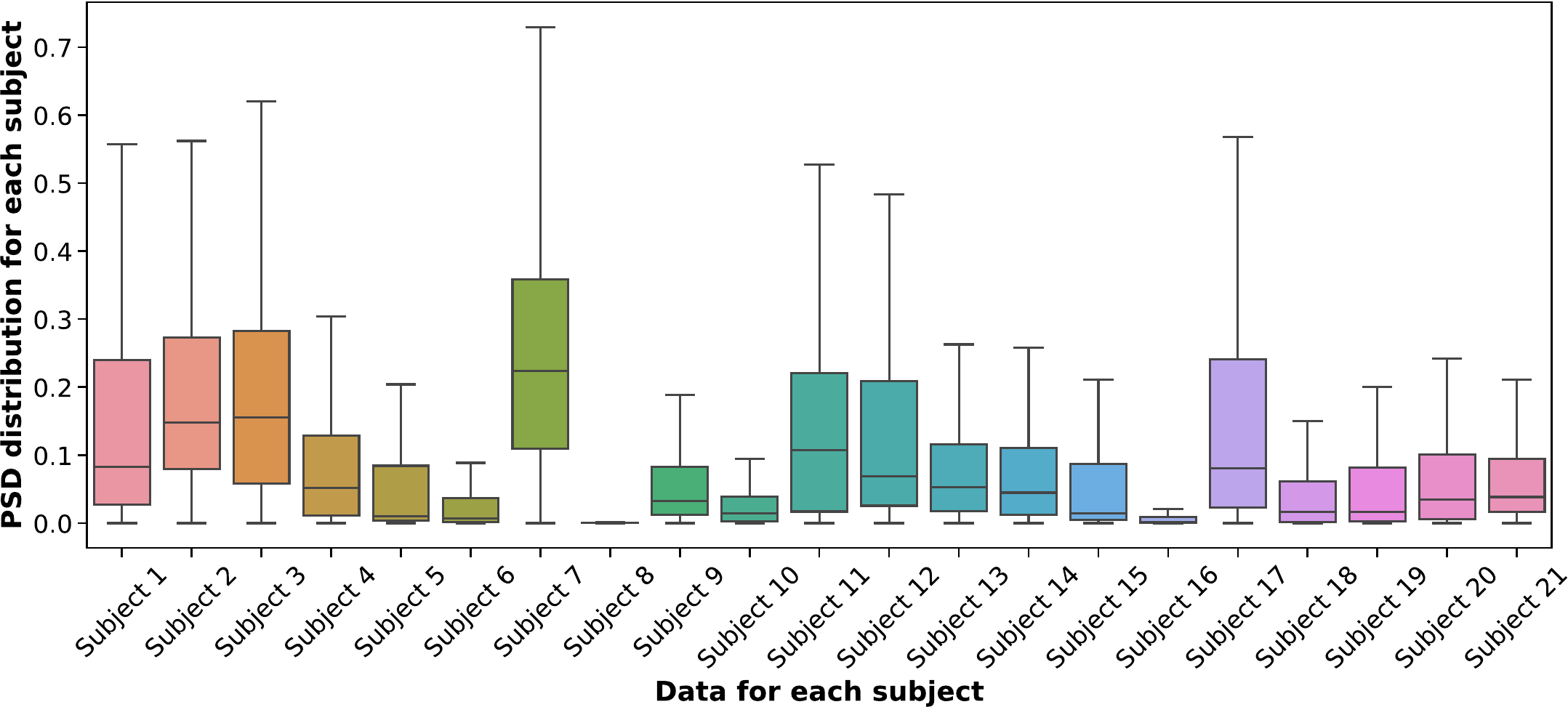}
    \caption{\addtxt{PSD distribution for each of the subjects included in the SEED-VIG dataset.}}
    \label{fig:Distribution-dataset}
\end{figure*}

\begin{table}[ht]
\centering

\caption{\addtxt{Example of the variability between subjects, presenting three configurations with different numbers of subjects usedin in the training set.}}
\label{tab:example-variability}

\resizebox{\columnwidth}{!}{
\begin{tabular}{lcccc} 
    \hline
    \addtxt{\textbf{Configuration}} & \addtxt{\textbf{Accuracy}} & \addtxt{\textbf{Precision}} & \addtxt{\textbf{Recall}} & \addtxt{\textbf{f1-score}} \\ 
    \hline
    
    \begin{tabular}{@{}l@{}}\addtxt{Train: Subject 1} \\ \addtxt{Test: Subject 21}\end{tabular} & \addtxt{0.35} & \addtxt{0.49} & \addtxt{0.46} & \addtxt{0.33} \\
    \hline

    \begin{tabular}{@{}l@{}}\addtxt{Train: Subjects 1-11} \\ \addtxt{Test: Subject 21}\end{tabular} & \addtxt{0.51} & \addtxt{0.50} & \addtxt{0.50} & \addtxt{0.46} \\
    \hline

    \begin{tabular}{@{}l@{}}\addtxt{Train: Subjects 1-19} \\ \addtxt{Test: Subject 21}\end{tabular} & \addtxt{0.60} & \addtxt{0.50} & \addtxt{0.54} & \addtxt{0.52} \\
    
    \hline
\end{tabular}}
\end{table}

\addtxt{It is also relevant to study the variability of the EEG data available in the SEED-VIG dataset for each subject. In particular, \figurename~\ref{fig:Distribution-dataset} depicts the PSD distribution for each of the 21 subjects in the dataset, highlighting a high inter-subject variability that could affect applying personalized models trained with data from one person to another user. To better study these variations, \tablename~\ref{tab:example-variability} presents three examples of models trained with different numbers of subjects and validated with data from Subject 21. Thus, training an individual model for the first subject and testing it on Subject 21 offered a 35\% accuracy. In contrast, a model trained with the first 11 subjects offered an accuracy of 51\% when evaluated with Subject 21. Finally, a model including the first 19 subjects resulted in a performance of 60\% accuracy when evaluated over the last subject. These results indicate that increasing the training set would improve the quality of the predictions for new subjects, thus justifying the need for scalable models.}

\subsection{Scalable Models}
Once the results of the individual models are available, the two best-performing algorithms in the individual approach (kNN and RF) and the most promising features for each one are selected for further study. In addition, the best combination for SVM is also included due to its large presence in the literature. These three combinations are used to evaluate \change{a single model}{further models}. Subsequently, each scalable model created is presented along with its performance.

\subsubsection{100 Models}
Regression performance is shown in \tablename~\ref{tab:m100-regression} where it can be seen that both kNN with PSD and RF with PSD+EOG have a fairly good RMSE and $R^2$. SVM with PSD, however, provides inferior performance compared to the other options. These results follow the same trend as the individual models presented in \tablename~\ref{tab:individual-regression}.

\begin{table}[ht]
\centering

\caption{Regression performance for the 100 models.}
\label{tab:m100-regression}

\resizebox{\columnwidth}{!}{
\begin{tabular}{lcC{2.5cm}cc} 
    \hline
    \textbf{Algorithm} & \textbf{Features} & \textbf{RMSE} & \textbf{\boldmath$R^2$} \\ 
    \hline
    SVM         & PSD               & $0.21$            & $0.43$ \\
    kNN         & PSD               & $0.14$            & $0.75$ \\
    \textbf{RF} & \textbf{PSD+EOG}  & \boldmath$0.12$   & \boldmath$0.80$ \\
    \hline
\end{tabular}}
\end{table}

As for the scalable classification models (\tablename~\ref{tab:m100-classification}), and in the same way as the regression models, the best options are again kNN and RF, in this case, both using PSD as features. Similarly to the results evaluating individual models, SVM has been the worst in performance.
\begin{table}[ht]
\centering

\caption{Classification performance for the 100 models.}
\label{tab:m100-classification}

\resizebox{\columnwidth}{!}{
\begin{tabular}{lcC{2cm}cccc} 
    \hline
    \textbf{Algorithm} & \textbf{Features} & \textbf{Accuracy} & \textbf{Precision} & \textbf{Recall} & \textbf{f1-score} \\ 
    \hline
    SVM         & PSD           & $0.63$ & $0.59$ & $0.51$ & $0.52$ \\
    kNN         & PSD           & $0.79$ & $0.76$ & $0.76$ & $0.76$ \\
    \textbf{RF} & \textbf{PSD}  & \boldmath$0.83$ & \boldmath$0.80$ & \boldmath$0.78$ & \boldmath$0.79$ \\
    \hline
\end{tabular}}
\end{table}

\subsubsection{90-10 Models}
Since there are a total of 23 experiments over 21 different subjects in the dataset, two subjects (avoiding those who had more than one experiment) corresponding to $\sim$10\% of the total are reserved for the evaluation of the model. Subsequently, 21 experiments from a total of 19 subjects are used for training the model\addtxt{s}.

As presented in \tablename~\ref{tab:9010-regression}, \change{SVM offers the worst performance of the three combinations studied. After that, kNN using PSD presents the best results, followed by RF using PSD and EOG data as the most promising combination.}{RF using PSD and EOG data is the most promising combination, followed by kNN using PSD. Finally, SVM offers the worst performance of the three combinations studied.}

\begin{table}[ht]
\centering

\caption{Regression performance for the 90-10 models.}
\label{tab:9010-regression}

\resizebox{\columnwidth}{!}{
\begin{tabular}{lcC{2.5cm}cc} 
    \hline
    \textbf{Algorithm} & \textbf{Features} & \textbf{RMSE} & \textbf{\boldmath$R^2$} \\ 
    \hline
    SVM         & PSD               & $0.26\pm0.06$ & $-1.91\pm0.10$ \\
    kNN         & PSD               & $0.20\pm0.08$ & $-0.66\pm0.04$ \\
    \textbf{RF} & \textbf{PSD+EOG}  & \boldmath$0.16\pm0.07$ & \boldmath$0.03\pm0.18$ \\
    \hline
    \end{tabular}}
\end{table}

Relative to the 90-10 classification models (see \tablename~\ref{tab:9010-classification}), it is important to note that, in this case, kNN with PSD as features performs slightly better than RF with PSD. \change{On the other hand}{In contrast}, SVM together with PSD offers results almost similar to the last two combinations mentioned, but always slightly worse.

\begin{table}[ht]
\centering

\caption{Classification performance for the 90-10 models.}
\label{tab:9010-classification}

\resizebox{\columnwidth}{!}{
\begin{tabular}{lcC{2cm}cccc} 
    \hline
    \textbf{Algorithm} & \textbf{Features} & \textbf{Accuracy} & \textbf{Precision} & \textbf{Recall} & \textbf{f1-score} \\ 
    \hline
    
    SVM & PSD & $0.57\pm0.23$ & $0.37\pm0.14$ & $0.41\pm0.18$ & $0.38\pm0.15$ \\
    
    \textbf{kNN} & \textbf{PSD} & \boldmath$0.60\pm0.17$ & \boldmath$0.46\pm0.12$ & \boldmath$0.48\pm0.12$ & \boldmath$0.46\pm0.12$ \\
    
    RF & PSD & $0.61\pm0.19$ & $0.44\pm0.18$ & $0.45\pm0.19$ & $0.43\pm0.17$ \\
    \hline
\end{tabular}}
\end{table}

\subsubsection{70-30 Models}
Analogous to the reasoning in the previous models, in this case, 16 experiments (from 14 different subjects) are assigned to model training while the remaining seven experiments, from seven different subjects, are reserved for evaluation.

\tablename~\ref{tab:7030-regression} presents the regression results while those corresponding to classification are shown in \tablename~\ref{tab:7030-classification}. In both approaches, SVM is always the worst of the three combinations. In regression, RF with PSD and EOG remains the best alternative, followed by kNN with PSD. Moving to classification, both kNN and RF with PSD are alternatives to consider, as RF offers a better accuracy compared to kNN but the second one slightly outperforms it in the rest of the metrics. It is interesting to mention that the average f1-score has fallen in all three cases below 40\%, which makes this set of models not as interesting as others presented above \addtxt{in terms of performance}.

\begin{table}[ht]
\centering

\caption{Regression performance for the 70-30 models.}
\label{tab:7030-regression}

\resizebox{\columnwidth}{!}{
\begin{tabular}{lcC{2.5cm}cc} 
    \hline
    \textbf{Algorithm} & \textbf{Features} & \textbf{RMSE} & \textbf{\boldmath$R^2$} \\ 
    \hline
    SVM & PSD & $0.26\pm0.07$ & $-1.90\pm3.02$ \\
    
    kNN & PSD & $0.22\pm0.05$ & $-0.70\pm0.65$ \\
    
    \textbf{RF} & \textbf{PSD+EOG} & \boldmath$0.18\pm0.05$ & \boldmath$-0.17\pm0.45$ \\
    
    \hline
\end{tabular}}
\end{table}

\begin{table}[ht]
\centering

\caption{Classification performance for the 70-30 models.}
\label{tab:7030-classification}

\resizebox{\columnwidth}{!}{
\begin{tabular}{lcC{2cm}cccc} 
    \hline
    \textbf{Algorithm} & \textbf{Features} & \textbf{Accuracy} & \textbf{Precision} & \textbf{Recall} & \textbf{f1-score} \\ 
    \hline
    
    SVM & PSD & $0.40\pm0.20$ & $0.38\pm0.07$ & $0.41\pm0.09$ & $0.29\pm0.14$ \\
    
    \textbf{kNN} & \textbf{PSD} & \boldmath$0.45\pm0.16$ & \boldmath$0.44\pm0.07$ & \boldmath$0.44\pm0.11$ & \boldmath$0.37\pm0.11$ \\
    
    \textbf{RF} & \textbf{PSD} & \boldmath$0.46\pm0.15$ & \boldmath$0.41\pm0.06$ & \boldmath$0.41\pm0.07$ & \boldmath$0.35\pm0.08$ \\
    
    \hline
\end{tabular}}
\end{table}

\subsection{Discussion}
The results for both individual and scalable models suggest that there are ML alternatives to SVM when estimating subjects' drowsiness, although the literature mainly use\addtxt{s} this algorithm, sometimes without exploring other \addtxt{ML} options. In the case of classification algorithms, to make a fair comparison with the literature, this section relies on accuracy and not on f1-score.

Regarding individual models, and comparing the metrics with \addtxt{the existing literature, Li et al.} \cite{Li:SEED-VIG-SVR:2018}, who used the same dataset but different features, the RMSE obtained in almost every combination of algorithm and features in the framework improves the RMSE of 0.17 provided in their research with SVR. Moreover, the accuracy of 93.6\% obtained \change{in}{by Zhu et al.} \cite{Zhu:PSD-CNN:2021} is close to the 86\% obtained by the best combination in the framework. \addtxt{Additionally, this work improves the results of many of the works studied from the literature using ML approaches).} 

\begin{table*}[!htb]
\centering
\caption{Comparison between the literature works reviewed and the results from the present work. \addtxt{The results are expressed in multiple metrics, where \textit{Acc} represents Accuracy, \textit{CC} means the Pearson correlation coefficient, \textit{RMSE} is the Root Mean Square Error, and \textit{$R^2$} indicates the coefficient of determination.}}
    \label{tab:discussion}
\resizebox{\textwidth}{!}{
\begin{tabular}{@{}llllll@{}}
\toprule
\textbf{Reference} & \textbf{Acquisition} & \textbf{Processing} & \textbf{Features} & \textbf{Classification} & \textbf{Results} \\ \midrule

\begin{tabular}{@{}c@{}}Chen et al. \\ (2015) \cite{Chen:DWT-SVM-ELM:2015}\end{tabular} & \addtxt{EEG, EOG} & \addtxt{N/A} & Four from DWT, blinking & \begin{tabular}{@{}l@{}}SVM \\ ELM \end{tabular} & \begin{tabular}{@{}l@{}}\addtxt{Acc: }96.90\% \\ \addtxt{Acc: }97.30\% \end{tabular} \\ \midrule

\begin{tabular}{@{}c@{}}Cheng et al. \\ (2018) \cite{Cheng:SVM-CNN:2018}\end{tabular} & \addtxt{EEG} & \addtxt{N/A} & PSD & \begin{tabular}{@{}l@{}}SVM \\ CNN \end{tabular} & \begin{tabular}{@{}l@{}}\addtxt{Acc: }64.05\% \\ \addtxt{Acc: }69.19\% \end{tabular} \\ \midrule

\begin{tabular}{@{}c@{}}Gwak et al. \\ (2018) \cite{Gwak:SVM-kNN-RF:2018}\end{tabular} & \addtxt{EEG, ECG, EOG} & \addtxt{Band-pass 1-40 Hz, ICA} & \begin{tabular}{@{}l@{}} PSD, ECG, EOG \\ simulation data \end{tabular} & \begin{tabular}{@{}l@{}}CNN \\ SVM \\ RF \end{tabular} & \begin{tabular}{@{}l@{}}\addtxt{Acc: }75.30\% \\ \addtxt{Acc: }78.60\% \\ \addtxt{Acc: }81.40\% \end{tabular} \\ \midrule

\begin{tabular}{@{}c@{}}Li et al. \\ (2018) \cite{Li:SEED-VIG-SVR:2018}\end{tabular} & \addtxt{EEG, EOG} & \addtxt{ICA, downsampling 125 Hz} & \begin{tabular}{@{}l@{}} 100 from EEG, \\ 36 from EOG \end{tabular} & \begin{tabular}{@{}l@{}}SVR \end{tabular} & \begin{tabular}{@{}l@{}}CC: 0.76 \\ RMSE: 0.17 \end{tabular} \\ \midrule

\begin{tabular}{@{}c@{}}Wei et al. \\ (2018) \cite{Wei:PSD-survey:2018}\end{tabular} & \addtxt{EEG} & \begin{tabular}{@{}c@{}}\addtxt{Band-pass 1-50 Hz, notch} \\ \addtxt{60 Hz, downsampling 250} \\ \addtxt{Hz, ASR}\end{tabular} & \begin{tabular}{@{}l@{}} PSD \end{tabular} & \begin{tabular}{@{}l@{}}kNN \\ LDA \\ SVM\end{tabular} & \begin{tabular}{@{}l@{}}\addtxt{Acc: }77.3\% \\ \addtxt{Acc: }79.4\% \\ \addtxt{Acc: }80.0\%\end{tabular} \\ \midrule

\begin{tabular}{@{}c@{}}Akbar and \\ Igasaki \\ (2019) \cite{Akbar:Recurrent-SVM:2019}\end{tabular} & \addtxt{EEG} & \addtxt{Band-pass 0.5-50 Hz} & \begin{tabular}{@{}l@{}} Hjorth parameters, PSD \end{tabular} & \begin{tabular}{@{}l@{}}SVM \\ Recurrent SVM \end{tabular} & \begin{tabular}{@{}l@{}} $R^2$: 0.64, RMSE: 0.56 \\ $R^2$: 0.83, RMSE: 0.15 \end{tabular} \\ \midrule

\begin{tabular}{@{}c@{}}Chakladar et \\ al. (2020) \cite{Chakladar:TD-SVM-RF:2020}\end{tabular} & \addtxt{EEG} & \addtxt{Band-pass 5-32 Hz} & \begin{tabular}{@{}c@{}}PSD, mean, SD, skewness, \\ kurtosis, AR, entropy\end{tabular} & \begin{tabular}{@{}l@{}}RF \\ SVM\end{tabular} & \begin{tabular}{@{}l@{}}\addtxt{Acc: }83.00\% \\ \addtxt{Acc: }83.33\% \end{tabular} \\ \midrule

\begin{tabular}{@{}c@{}}Cui et al. \\ (2021) \cite{Cui:Oz-survey:2021}\end{tabular} & \addtxt{EEG} & \begin{tabular}{@{}c@{}}\addtxt{Band-pass 1-50 Hz, AAR,} \\ \addtxt{downsampling 128 Hz}\end{tabular} & \begin{tabular}{@{}c@{}}Oz channel\end{tabular} & \begin{tabular}{@{}l@{}}DT \\ RF \\ kNN \\ Gaussian Naïve Bayes \\ SVM \end{tabular} & \begin{tabular}{@{}l@{}} \addtxt{Acc:} 60.70\% \\ \addtxt{Acc: }62.60\% \\ \addtxt{Acc: }63.42\% \\ \addtxt{Acc: }67.44\% \\ \addtxt{Acc: }69.72\% \end{tabular} \\ \midrule

\begin{tabular}{@{}c@{}}\addtxt{Paulo et al.} \\ \addtxt{(2021) \cite{Paulo:DL:2021}}\end{tabular} & \addtxt{EEG} & \begin{tabular}{@{}l@{}}\addtxt{Band-pass 1-50 Hz, AAR}\end{tabular} & \addtxt{Raw EEG} & \begin{tabular}{@{}l@{}}\addtxt{CNN} \end{tabular} & \begin{tabular}{@{}l@{}}\addtxt{Acc: 75.87\%} \end{tabular} \\ \midrule

\begin{tabular}{@{}c@{}}Shen et al. \\ (2021) \cite{Shen:TN:2021}\end{tabular} & \addtxt{EEG} & \begin{tabular}{@{}c@{}}\addtxt{Band-pass 1-50 Hz, AAR,} \\ \addtxt{downsampling 120 Hz}\end{tabular} & PSD & \begin{tabular}{@{}l@{}}SVM \end{tabular} & \begin{tabular}{@{}l@{}}\addtxt{Acc: }62.51\% \end{tabular} \\ \midrule

\begin{tabular}{@{}c@{}}\addtxt{Qian et al.} \\ \addtxt{(2021) \cite{Qian:Drowsiness:2021}}\end{tabular} & \addtxt{EEG} & \begin{tabular}{@{}c@{}}\addtxt{Band-pass $<$30 Hz}\end{tabular} & \addtxt{PSD} & \begin{tabular}{@{}l@{}}\addtxt{BCDC} \\ \addtxt{GSVM} \end{tabular} & \begin{tabular}{@{}l@{}}\addtxt{Acc: 94.3\%} \\ \addtxt{Acc: 93.7\%} \end{tabular} \\ \midrule

\begin{tabular}{@{}c@{}}Zhu et al. \\ (2021) \cite{Zhu:PSD-CNN:2021}\end{tabular} & \addtxt{EEG} & \begin{tabular}{@{}c@{}}\addtxt{Band-pass 1-60 Hz, notch} \\ \addtxt{50 Hz, ICA, wavelet threshold}\end{tabular} & Raw EEG & \begin{tabular}{@{}l@{}}CNN-Inception \\ CNN-AlexNet \end{tabular} & \begin{tabular}{@{}l@{}}\addtxt{Acc: }93.6\% \\ \addtxt{Acc: }94.68\% \end{tabular} \\ \midrule

\begin{tabular}{@{}c@{}}\addtxt{Arefnezhad et} \\ \addtxt{al. (2022)} \cite{Arefnezhad:Encoding:2022}\end{tabular} & \begin{tabular}{@{}c@{}}\addtxt{EEG}\end{tabular} & \addtxt{ICA} & \begin{tabular}{@{}c@{}}\addtxt{Skewness alpha, delta power,} \\ \addtxt{theta power, Hjorth delta mobility} \end{tabular} & \begin{tabular}{@{}l@{}}\addtxt{Encoder-decoder} \end{tabular} & \begin{tabular}{@{}l@{}}\addtxt{RMSE: 0.117} \end{tabular} \\ \midrule

\begin{tabular}{@{}c@{}}\addtxt{Arif et al.} \\ \addtxt{(2023)} \cite{Arif:ML:2023}\end{tabular} & \begin{tabular}{@{}c@{}}\addtxt{EEG}\end{tabular} & \begin{tabular}{@{}c@{}}\addtxt{Band-pass} \\ \addtxt{0.5-40Hz, Notch} \end{tabular} & \begin{tabular}{@{}c@{}}\addtxt{PSD, band power ratio indices} \end{tabular} & \begin{tabular}{@{}l@{}} \addtxt{Discriminant analysis} \\ \addtxt{LR} \\ \addtxt{Naïve Bayes} \\ \addtxt{SVM} \\ \addtxt{DT} \\ \addtxt{kNN} \\ \addtxt{Ensemble classier} \end{tabular} & \begin{tabular}{@{}l@{}}\addtxt{Acc: 63.5\%} \\ \addtxt{Acc: 63.6\%} \\ \addtxt{Acc: 67.4\%} \\ \addtxt{Acc: 75.7\%} \\ \addtxt{Acc: 77.4\%} \\ \addtxt{Acc: 78.5\%} \\ \addtxt{Acc: 85.6\%}\end{tabular} \\ \midrule

\begin{tabular}{@{}c@{}}This work\end{tabular} & \begin{tabular}{@{}c@{}}EEG, EOG\end{tabular} & \begin{tabular}{@{}c@{}}\addtxt{Notch 60 Hz, band-pass} \\ \addtxt{1-30Hz, downsampling} \\ \addtxt{60 Hz, ICA} \end{tabular} & \begin{tabular}{@{}c@{}}\addtxt{EEG, PSD, EOG} \end{tabular} & \begin{tabular}{@{}l@{}} Gaussian Process \\ SVM \\ Decision Trees \\ kNN \\ Random Forest \end{tabular} & \begin{tabular}{@{}l@{}}\addtxt{Acc: 71\%} \\ \addtxt{Acc: 76\%} \\ \addtxt{Acc: 80\%} \\ \addtxt{Acc: 85\%} \\ \addtxt{Acc: 86\%} \end{tabular} \\

\bottomrule
\end{tabular}}
\end{table*}

As can be seen in \tablename~\ref{tab:discussion}, the best results for the trained individual models are in line with the claims of Gwak et al. \rmvtxt{(2018)} \cite{Gwak:SVM-kNN-RF:2018} where RF performed better than SVM. \change{On the contrary}{However}, the results contradict Cui et al. \rmvtxt{(2021)} \cite{Cui:Oz-survey:2021} and Chakladar et al. \rmvtxt{(2020)} \cite{Chakladar:TD-SVM-RF:2020} since in both studies, SVM performed better or similarly than the other tested \addtxt{ML-based} algorithms. This, which may look controversial, can be explained by the features employed by Gwak et al. \rmvtxt{(2018)} \cite{Gwak:SVM-kNN-RF:2018} and the present study, where PSD and EOG features are used. \rmvtxt{On the contrary,} Cui et al. \rmvtxt{(2021)} \cite{Cui:Oz-survey:2021} used an entire EEG channel as feature and Chakladar et al. \rmvtxt{(2020)} \cite{Chakladar:TD-SVM-RF:2020} combined PSD with time-domain features. 

Therefore, a common pattern is observed: if PSD is used, the model performance obtained is increased compared to not using it and, thus, algorithms such as RF tend to perform better or, at least, similar to SVM. This pattern is \change{also found}{observed} in studies like \addtxt{Gwak et al.} \cite{Gwak:SVM-kNN-RF:2018} with an accuracy of 81.40\% using RF and \addtxt{Chakladar et al.} \cite{Chakladar:TD-SVM-RF:2020} with 83.33\% in SVM and 83.00\% in RF. This may contribute to a clearer understanding of which features and algorithms should be taken into account when considering training a model for the prediction of drowsiness while driving.

\addtxt{The relevance of PSD as a feature can be clearly explained by how the different EEG brainwaves change between cognitive states. In particular, beta and gamma waves are predominant in demanding cognitive states, such as problem-solving, focused attention, or information processing. In contrast, during drowsy states, theta and alpha are the most common waves. In particular, theta waves are related to relaxation, drowsiness, and early stages of sleep, while alpha activity is predominant when subjects are awake but relaxed \cite{Ramadan2021}. Focusing on the present work, Zheng et al. \cite{seed-vig-dataset} published detailed information regarding the SEED-VIG dataset, corroborating that the employed dataset predominates theta and alpha brain waves during drowsy states. This situation is also contrasted by the importance of the features of RF used in the present study, where theta, alpha, and gamma frequency bands were the most representative in all models studied.}

Concerning scalable models, the 100 model performance is similar to the individual models, implying that having just one model for all users could be enough, compared to having one model per subject. Moreover, the 90-10 and 70-30 models reach an accuracy of $0.60\pm0.17$ and $0.46\pm0.15$\addtxt{,} respectively. In both cases, the performance is greater than 33\%, which represents the accuracy of predicting the level of sleepiness randomly. Because of that, these results suggest that it could be possible to develop a scalable model which can predict drowsiness in subjects that are not involved in the experimentation and training phase of the model, although this may depend on the similarity of the subject's features distribution to those used during training.

\change{Nevertheless, the generalization of the results is limited by the data provided by the dataset as the models are trained for a specific group of 21 subjects who participated in its creation}{Despite the promising results obtained, this research presents certain limitations. First, the results depend on the amount and quality of the data used. In particular, the models were trained with a specific group of 21 subjects, which could be insufficient to reach a substantial model generalization. Secondly, this research is limited to exploring the performance of ML algorithms. It is convenient to have access to a larger dataset to explore more complex models, such as those based on DL, able to detect more sophisticated patterns and, thus, achieve better performance.}. \change{Further}{Based on these limitations, further} research is needed to establish the generalization of \change{our}{the} findings, using a larger number of subjects during the training \addtxt{and testing} phase\addtxt{s}.

\section{Conclusions}\label{sec:5_conclusions}
Drowsiness while driving is a major source of accidents and fatalities. To try to improve this situation, this research presents a framework for drowsiness detection in driving scenarios employing BCIs based on EEG, where different algorithms and feature vectors are used for regression and three-class classification. This is done for both individual and scalable models, where the first ones offer predictions for just one subject, whereas the latter are capable of estimating sleepiness in various subjects despite not having been trained with data from them. In particular, three configurations of scalable models are evaluated, based on the percentage of users \change{used}{employed} to evaluate the models that are not included in the training phase. To \change{evaluate}{validate} the framework, the SEED-VIG dataset is used, which contains a total of 23 experiments performed in a driving simulator involving 21 different subjects. The labels to be predicted are PERCLOS values whose discretization is obtained via a dynamic PERCLOS discretization algorithm, taking into account the physiological particularities of each subject. 

The results obtained suggest that PSD features are highly relevant when estimating drowsiness since the best performance for almost every tested algorithm involved PSD, regardless of the learning technique or type of model used. Also, this research illustrates that algorithms such as kNN, RF, or DT may perform equal or better than SVM, the most used algorithm in the literature. Furthermore, GP algorithms are the worst in performance, due to the intrinsic properties of the EEG signals. Lastly, looking at the drowsiness detection performance of the different trained models, the individual models offer the best results, with the limitation of being restricted to a single subject, not being scalable and valid for new \change{subjects}{users}. Next, 100 models, which use the 100\% of the subjects for training and testing, provide remarkably similar results to the previous ones while reducing the complexity of the experimentation into a single \addtxt{general} model. Finally, the performance of 90-10 and 70-30 models, which reserve the 10\% and 30\% of subjects for evaluating the models, respectively, show the possibility of predicting drowsiness in subjects not involved during the training phase of the model, although they present a degradation in performance.

As future work, this study first proposes the generation of a \addtxt{new} dataset using a BCI \rmvtxt{of our own}, aiming to compare the current results with those obtained \change{with this equipment}{from using a larger dataset}. Next, it is intended to apply deep learning algorithms for drowsiness estimation, as they are becoming increasingly popular in the literature and could provide better results. Lastly, it is planned to continue working with the scalable 90-10 and 70-30 models to obtain more realistic and robust models capable of predicting drowsiness on a larger set of new subjects.

\backmatter

\bmhead{Acknowledgments}
This work has been partially supported by (a) 21628/FPI/21 and 21629/FPI/21 Fundación Séneca, cofunded by Bit \& Brain Technologies S.L. Región de Murcia (Spain), (b) Bit \& Brain Technologies under the project CyberBrain: Cybersecurity in BCI for Advanced Driver Assistance, associated with the University of Murcia (Spain), (c) the Swiss Federal Office for Defense Procurement (armasuisse) with the CyberTracer (CYD-C-2020003) project, and (d) the University of Zürich UZH.

\section*{Declarations}
\begin{itemize}
    \item Competing interests: Not applicable.
    \item Availability of data and materials: The datasets generated during and/or analyzed during the current study are available from the corresponding author on reasonable request.
\end{itemize}

\bibliographystyle{sn-basic}
\bibliography{sn-bibliography}


\end{document}